\documentclass[aps,prb,twocolumn,nopacs,superscriptaddress,longbibliography]{revtex4-1}
\usepackage[breaklinks=true,colorlinks,citecolor=blue,linkcolor=blue,urlcolor=blue]{hyperref}
\usepackage{graphicx,dcolumn,bm,multirow,hyperref,mathtools,braket,color,url,epstopdf,amssymb,amsmath}
\usepackage{xcolor}
\usepackage{graphicx}
\usepackage{dcolumn}
\usepackage{bm}

\usepackage{ulem}

\usepackage{multirow}

\usepackage[mathscr]{euscript}

\DeclareSymbolFont{rsfs}{U}{rsfs}{m}{n}
\DeclareSymbolFontAlphabet{\mathscrsfs}{rsfs}

\begin{document}

\preprint{APS/123-QED}


\title{Shubnikov-de Haas and de Haas-van Alphen oscillation in  \\Czochralski grown CoSi single crystal}

\author{Souvik Sasmal}
\email{sasmalsouvik6@gmail.com}
\affiliation{Department of Condensed Matter Physics and Materials Science, Tata Institute of Fundamental Research, Homi Bhabha Road, Colaba, Mumbai 400 005, India.}

\author{Gourav Dwari}
\affiliation{Department of Condensed Matter Physics and Materials Science, Tata Institute of Fundamental Research, Homi Bhabha Road, Colaba, Mumbai 400 005, India.}

\author{Bishal Baran Maity}
\affiliation{Department of Condensed Matter Physics and Materials Science, Tata Institute of Fundamental Research, Homi Bhabha Road, Colaba, Mumbai 400 005, India.}

\author{Vikas Saini}
\affiliation{Department of Condensed Matter Physics and Materials Science, Tata Institute of Fundamental Research, Homi Bhabha Road, Colaba, Mumbai 400 005, India.}

\author{Rajib Mondal}
\affiliation{UGC-DAE Consortium for Scientific Research, Kolkata Centre, Bidhannagar, Kolkata 700 106, India.}


\author{A. Thamizhavel}
\email{thamizh@tifr.res.in}
\affiliation{Department of Condensed Matter Physics and Materials Science, Tata Institute of Fundamental Research, Homi Bhabha Road, Colaba, Mumbai 400 005, India.}


\begin{abstract}
Anisotropic transport, Shubnikov-de Haas (SdH), and de Haas-van Alphen (dHvA) quantum oscillations studies are reported on a high-quality CoSi single crystal grown by the Czochralski method. Temperature-dependent resistivities indicate the dominating electron-electron scattering. Magnetoresistance (MR) at 2~K reaches 610\% for $I~\parallel~[111]$ and $B~\parallel~$[01\={1}], whereas it is 500\% for $I~\parallel~$[01\={1}] and $B~\parallel~[111]$. A negative slope in field-dependent Hall resistivity suggests electrons are the majority carriers. The carrier concentration extracted from Hall conductivity indicates no electron-hole compensation. In $3D$ CoSi, the electron transport lifetime is found to be approximately in the same order as quantum lifetime, whereas in $2D$ electron gas the long-range scattering drives the transport life much larger than the quantum lifetime. From linear and Hall SdH oscillations the effective masses and Dingle temperatures have been calculated. The dHvA oscillation reveals three  frequencies at 18~($\gamma$), 558 ($\alpha$) and 663~T ($\beta$)), whereas, SdH oscillation results in  only two frequencies $\alpha$ and $\beta$. The $\gamma$ frequency observed in dHvA oscillation is a tiny hole pocket at the $\Gamma$ point.

\end{abstract}

\maketitle


\section{Introduction}

 The recent theoretical and experimental studies on CoSi compound crystallizing in the $B20$-type cubic noncentrosymmetric crystal structure has attracted a renewed interest to study the transport and Fermi surface properties as it hosts new fermionic excitations~\cite{PhysRevLett.122.076402,Rao2019,Yuaneaaw9485,Sanchez2019,PhysRevLett.119.206402}. CoSi hosts two types of chiral topological fermions, a spin-1 chiral fermion at the center and a double Weyl fermion in the boundary of the Brillouin zone. Furthermore, it exhibits the Fermi arc state on (001) and (011) surfaces which host unconventional chiral fermions, whereas no such surface states on (111) surface~\cite{Yuaneaaw9485,Li2019,Sanchez2019} were observed. Additionally, in CoSi, the bulk charge carriers possess nontrivial topology which results in Fermi arc surface state on the side surface~\cite{PhysRevLett.119.206402}. The behavior of such a new type of fermions is quite intriguing. To explore the robust and unusual behavior of the charge carriers, a thorough investigation of CoSi is warranted on a pure single crystal. In presence of high magnetic field, both the linear and Hall resistivity tensors oscillate periodically with inverse magnetic field and the periodicity is determined by the carrier density~\cite{shoenberg2009magnetic,PhysRevB.39.1120}.   The theoretical treatments are given by Ando~\cite{doi:10.1143/JPSJ.37.1233}, Ando et.al.,~\cite{doi:10.1143/JPSJ.39.279} and Matsamoto et.al.,~\cite{Isihara_1986} in which they considered only one type of carrier lifetime for both linear and Hall. But many experimental results revealed the transport lifetime ($\tau_p$), which involves scattering angle, is different from the quantum lifetime ($\tau_q$) which is given by the total scattering rate. Typically, for $2D$ electron gas in GaAs/Ga$_{1-x}$Al$_x$As heterostructure, associated with long-range scattering potential, these two scattering lifetimes differ by  $10-100$ times~\cite{PhysRevB.32.8126, PhysRevB.32.8442,PhysRevLett.51.2226,PhysRevB.44.3793}. In Cd$_3$As$_2$ the strong suppression of backscattering results in an ultra-high mobility and $(\tau_p/\tau_q)$ ratio is $\sim 10^{4}$~~\cite{Liang2015}, whereas this ratio is 5000 for WP$_2$~\cite{Kumar2017}.  Since CoSi hosts a new types of fermions we wanted to investigate the $\tau_p$ and $\tau_q$.  To accomplish this we have grown a single crystal of CoSi by the Czochralski method in a tetra-arc furnace.  One of the advantages of Czochralski method is that the crystal is grown from a stoichiometric melt without any addition of a third element like flux or transporting agent.  In this manuscript, we present the results on electrical transport, Hall resistivity, and quantum oscillations of a pure CoSi single crystal. 

The temperature-dependent resistivity measurements show electron-electron interaction is dominant. From the  Hall conductivity, the electron and hole concentrations are calculated and  no sign of compensation is observed. The high magnetic field data show SdH quantum oscillation in  linear resistivity.  Incidentally, the Hall resistivity also revealed Shubnikov de Haas (SdH) oscillations at high magnetic fields indicating a good quality of the grown single crystal.  No such SdH oscillation in Hall resistivity has been observed in previously reported CoSi single crystal.  The two frequencies $\alpha$ and $\beta$ are observed in the SdH oscillation in the previous reports on CoSi single crystal~\cite{PhysRevB.100.045104, CoSi2019}.  We have also observed these two frequencies $\alpha$ and $\beta$ at 558 and 663~T respectively from the SdH oscillation, which correspond to the electron pockets at $R$ point in the Brillouin zone (BZ). Magnetization data also show quantum oscillations, indicating the de Haas-van Alphen (dHvA) oscillation. For both $B~\parallel~[111]$ and [01\={1}] directions, an additional frequency at 18~T ($\gamma$) is observed at $2$~K apart from the two frequencies ($\alpha$ and $\beta$) observed in SdH oscillations. Though this dhVA oscillation frequency $\gamma$ corresponding to the tiny hole pocket is observed in Ref.~\cite{PhysRevB.102.115129}, the Dingle temperature, quantum lifetime and Berry phase have not been calculated. Here we elucidate the effective mass and Dingle temperature of the $\gamma$ pocket which is found to be small compared to $\alpha$ and $\beta$.

\section{Methods}

\begin{figure}[t]
\centering
\includegraphics[width=0.45\textwidth]{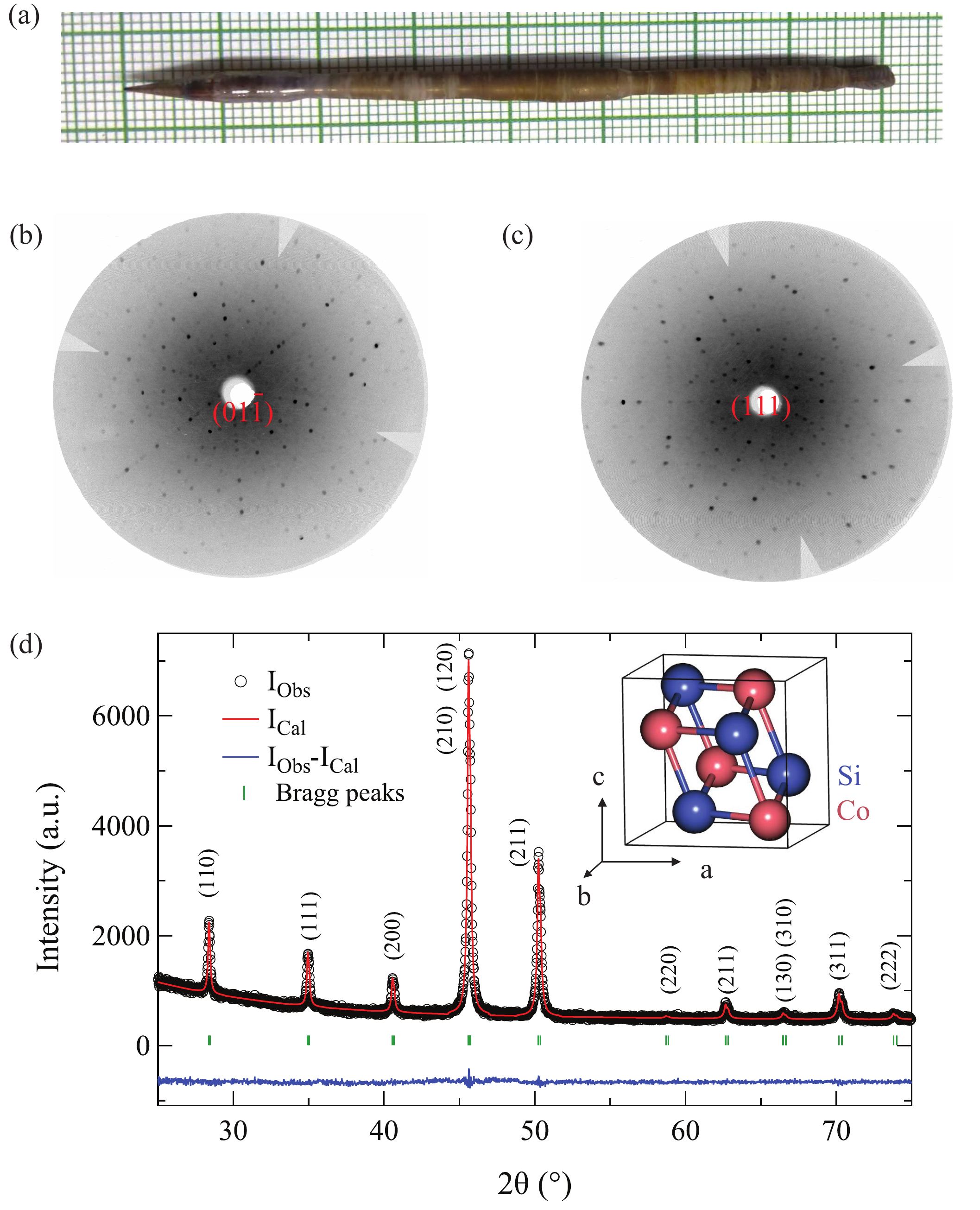}
\caption{(a) An image of grown single crystal of CoSi by Czochralski method. (b),(c) Laue diffraction pattern of crystallographic plane (01\={1}) and (111). (d) Powder XRD pattern at 300~K along with the Rietveld refinement. }
\label{crystal}
\end{figure}

The binary CoSi compound melts congruently at 1455~$^{\circ}$C and hence a single crystal of CoSi has been grown by the Cozchralski method in a tetra-arc furnace (TECHNOSEARCH CORP., Japan).  High purity starting elements of Co and Si were weighed in the molar ratio of 1:1.  A total charge of about 10~g has been remelted several times and a polycrystalline seed was cut from the ingot.  The polycrystalline seed was carefully inserted into the molten solution of CoSi and pulled rapidly at the rate of  50~mm/h.  Once the steady-state condition was achieved, the crystal pulling rate was maintained at 10~mm/h.  Roughly $8$~cm long single crystal with a typical diameter of $3-4$~mm has been grown.   The as grown single crystal and the Laue diffraction corresponding to (011) and (111) planes are shown in Fig.~\ref{crystal}(a)-(c). CoSi crystallizes in cubic structure with space group P2$_1$3 (No.~198).  Figure~\ref{crystal}(d) depicts the room temperature powder XRD along with the Rietveld refinement. It is evident that there is no sign of impurity peaks suggesting that the grown crystal is phase pure. From the powder XRD, the obtained lattice parameter is $a = 4.444(1)$.  Electrical transport measurements have been performed on a well aligned bar-shaped sample of dimension $2\times1.5\times0.3$~mm$^3$ in a Quantum Design Physical Property Measurement System (PPMS) in the temperature range $2-300$~K and magnetic fields up to $14$~T.

\section{Results}
\subsection{Eletrical Transport Studies}

\begin{figure}[]
\centering
\includegraphics[width=0.48\textwidth]{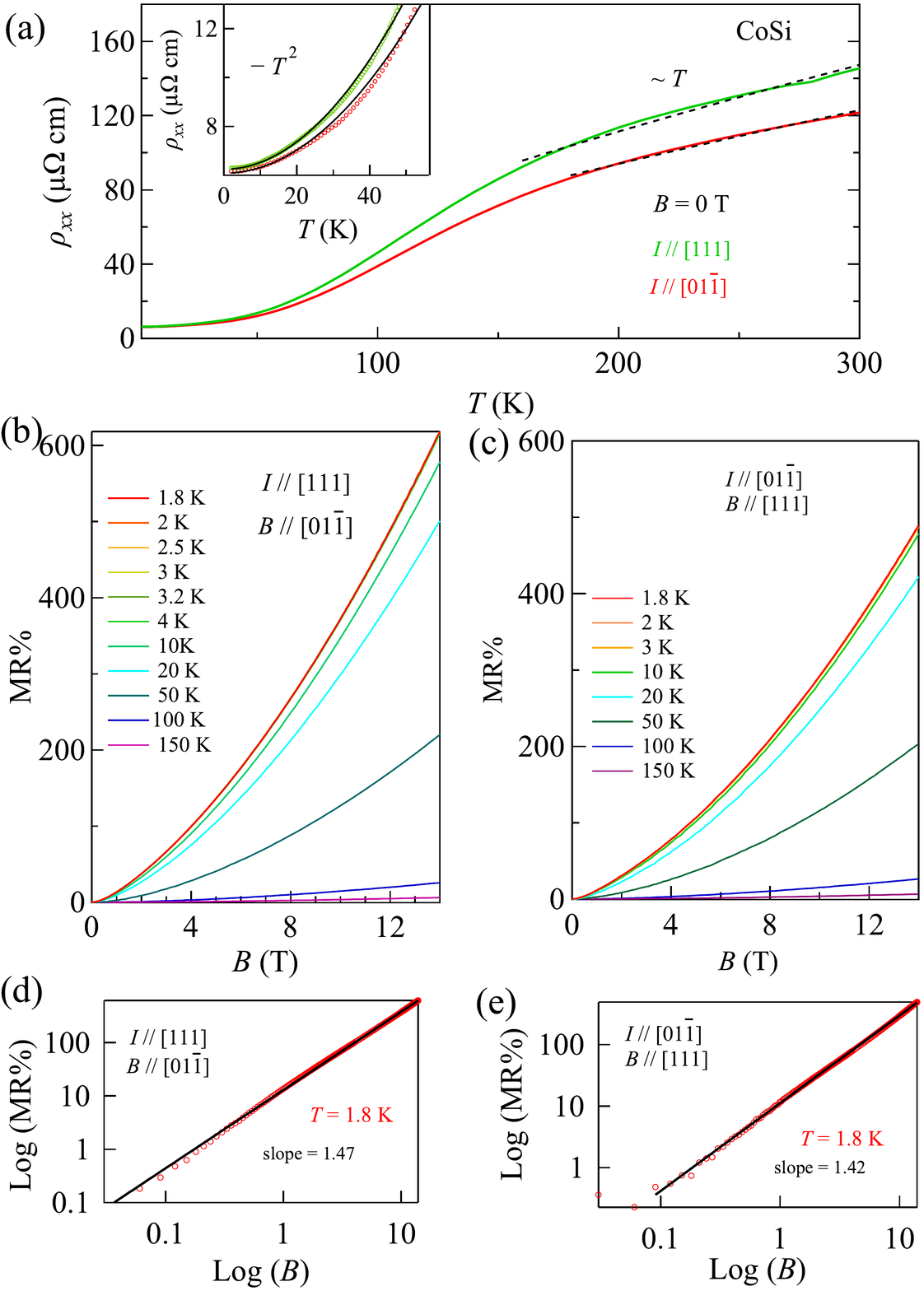}
\caption{(a) Temperature-dependent linear resistivity ($\rho_{xx}$) at zero magnetic field. High temperature resistivity  follows linear behavior ($\sim~T$). Inset: low temperature resistivity follows quadratic ($\sim T^2$) nature. (b),(c) MRs at different temperature for $I~\parallel~[111]$ and $I~\parallel~$[01\={1}], respectively. (d),(e) Log(MR\%) $vs$ Log($B$) plots are linear, with slopes of $1.47$ for $I~\parallel~[111]$ and $1.42$ for $I~\parallel~$[01\={1}].}
\label{RT}
\end{figure}

All electrical transport and magnetic measurements have been carried out on  bar-shaped samples cut along particular crystallographic directions. Figure.~\ref{RT}(a) shows temperature ($T$) dependent resistivity for different current ($I$) directions. The residual resistivity ratio (RRR) $\frac{\rho_{xx} (300 ~\rm{K})}{\rho_{xx} (2~\rm{K})}$ is 25 for $I~\parallel~[111]$ and 20 for $I~\parallel$~[01\={1}]. These  RRR values are higher than previously reported single crystal~\cite{PhysRevB.102.115129,CoSi2019}, and such high RRR values indicate the grown crystal is of high quality. In the high-temperature region for both $I~\parallel~[111]$ and $I~\parallel$~[01\={1}], the resistivity follows almost linear temperature dependence ($\sim~T$) indicating a dominant electron-phonon interaction whereas at lower temperature it follows as a quadratic dependence ($\sim~T^2$) indicating a strong electron-electron interaction. Typically, in a normal metal, at a higher temperature, all phonons modes are populated, and the lattice vibration largely contributes electron scattering to electrical resistivity, and the lattice resistivity is proportional to absolute temperature ($\sim~T$). At lower temperatures, if electron-electron scattering is ignored, it has different characteristics due to the correlation of motions of ions on neighboring lattice sites, so that the resistivity falls off rapidly with the fifth power of the resistivity ($\sim~T^5$)~\cite{https://doi.org/10.1002/andp.19334080504,PhysRevLett.105.256805,abrikosov2017fundamentals,Ziman2001}. But when the electron-electron scattering is much prominent at a lower temperature the resistivity follows square of temperature ($\sim~T^2$)~\cite{Ziman2001}.  Inset in Fig.~\ref{RT}(a) reveals the electron-electron interaction is much stronger in the low-temperature regime. Fig.~\ref{RT}(b) and (c) show magnetoresistances (MRs) at different temperatures for $I~\parallel$~[111] and [01\={1}] directions, respectively. At 1.8~K, the MRs reach about $\sim$~610\% for $I~\parallel~[111]$ and $\sim$~500\% for $I~\parallel$~[01\={1}]. As temperature increases the MRs decrease due to larger a phononic contribution. In case of electron hole compensation, the MR typically follows a $B^2$ behavior~\cite{Tafti2016,PhysRevB.94.241119,PhysRevB.102.115158,PhysRevB.97.205130}. In Fig.~\ref{RT}(d) and (e), the MRs follow $\sim~B^{1.47}$ and $\sim~B^{1.42}$ for $I~\parallel~[111]$ and $I~\parallel~$[01\={1}] at 1.8~K, respectively, which  suggest that CoSi is not a compensated system.

\begin{figure}[]
\centering
\includegraphics[width=0.48\textwidth]{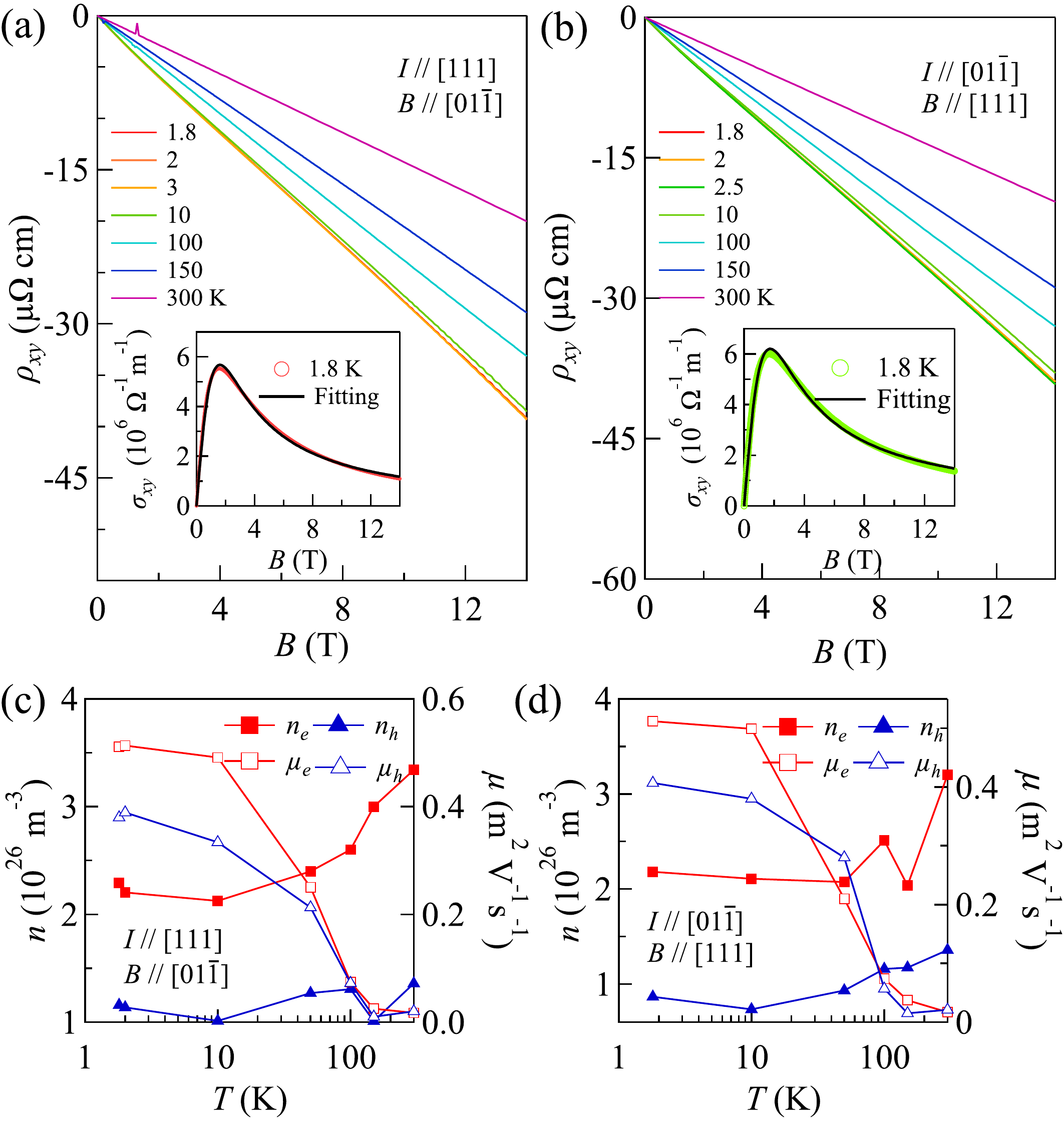}
\caption{(a),(b) Hall resistivity plots as a function of magnetic field at different temperatures for $I~\parallel~[111]$ and $I~\parallel~$[01\={1}], respectively. Inset: Hall conductivity fitting using two-band model at 1.8~K. (c),(d) Both electron and Hole concentrations and mobilities at different temperature for different current directions.}
\label{Hall}
\end{figure}

Hall resistivity plots ($\rho_{xy}$) are shown in Fig.~\ref{Hall}(a) and (b). Here, we have antisymmetrized the Hall resistivity $\rho_{xy} = [\rho_{xy}^{raw}(+B)-\rho_{xy}^{raw}(-B)]/2$.  $\rho_{xy}$ is negative suggesting electrons are the dominant carriers in this system. Using a two-band model electron and Hole carrier concentration and mobilities are calculated. According to the Drude model, the Hall conductivity ($\sigma_{xy}$) can be considered as the sum  of contributions from both electrons and holes:

\begin{equation}
\label{eq1}
    \sigma_{xy} = \frac{en_h\mu_h^2B}{1+(\mu_hB)^2}-\frac{en_e\mu_e^2B}{1+(\mu_eB)^2}
\end{equation}

Here $\sigma_{xy} = -\frac{\rho_{xy}}{\rho_{xy}^2+\rho_{xx}^2}$, $e$ is the electronic charge $1.62\times10^{-19}~\rm{C}$. $n_h$ and $\mu_h$ are carrier concentration and mobility of hole, whereas $n_e$ and $\mu_e$ are carrier concentration and mobility of electron. Inset of Fig.~\ref{Hall}(a) and (b) show a good fit of the two-band model at 1.8~K. Fig.~\ref{Hall}(c) and (d) show temperature-dependent electron and hole concentrations and mobilities. There are a subtle change in the carrier concentrations as the temperature increases, however, the mobilities decrease with an increase in temperature.  The values of carrier concentrations and mobilities are similar for both the current and field directions. At $2$~K, for $I~\parallel~[111]$ and $B~\parallel~$[01\={1}], the electron concentration ($n_e$) is $2.3\times10^{26}$~m$^{-3}$ and hole concentration ($n_h$) is $1.2\times10^{26}$~m$^{-3}$, which are in good agreement with the previously published data~\cite{PhysRevB.100.045104}. Here the ratio $n_e/n_h$ is $\approx 2$, which suggests that the carriers are not compensated. The electron and hole mobilities are estimated as $\mu_e = 0.51$~m$^2$V$^{-1}$s$^{-1}$ and $\mu_h = 0.4$~m$^2$V$^{-1}$s$^{-1}$, respectively at $T=2$~K.  The transport life time ($\tau_p$) can be calculated from the relation: $\tau_p = \frac{\mu~m^*}{e}$. Here, $e$ is the electron charge, $m^*$ is effective mass and $\mu$ is carrier mobility. For $I~\parallel~[111]$, The estimated electron (hole) transport lifetime $\tau_{p_e}~(\tau_{p_h})$ is $2.5\times10^{-12}$~s ($3.5\times10^{-13}$~s), we have used the effective mass values obtained from the quantum oscillation analysis described in the next section.   Similarly,  quantum lifetimes of electron and hole are extracted from SdH and dHvA oscillations (discussed later). It is found that both the transport and quantum lifetime are in the same order. It is to be mentioned here that in $2D$ GaAs/Ga$_{1-x}$Al$_x$As or in Cd$_3$As$_2$ $3D$ bulk systems a much larger $\tau_{\rm p}$ is observed compared to $\tau_{\rm q}$ owing to the suppression of the back-scattering of the charge carriers.  Usually, the transport life time is weighted by scattering angle. When the charge carrier drive in long-range potential scattering, small-angle scattering is dominated and the $\tau_p$ becomes large~\cite{PhysRevB.32.8126,PhysRevB.39.1120}. Subsequently, very high mobilities $3\times10^{7}~\rm{cm}^{2}\rm{V}^{-1}\rm{s}^{-1}$ for GaAs - based 2DEG and $9\times10^{6}~\rm{cm}^{2}\rm{V}^{-1}\rm{s}^{-1}$ for Cd$_3$As$_2$ are observed~\cite{Schlom2010,Liang2015}. In the case of CoSi, the mobility is not very high and causes such low transport to quantum life time ratio ($\tau_p/\tau_q$). Furthermore, electron-electron interaction is present in the system. Although, the large or small $\tau_p/\tau_q$ does not necessarily indicate whether the scattering is long or short-range, in general, if $\tau_p/\tau_q$ is small it is identical to that of the short-range scattering potential.   Thus it speculates the short-range scattering in the CoSi. This short-range scattering can be associated with point defect scattering, and the strength of the potential can be understood from low temperature electrical transport as mentioned by Pshenay-Sever \textit{et al}~\cite{Pshenay2018}. In Co$_x$Fe$_{1-x}$Si~\cite{Pshenay2018}, this point defect scattering prevails below 50~K, whereas at the higher temperatures the scattering is due to electron-phonon and point defect scattering contributed by the inter and intra band scattering.  Similarly, here we observed that electron-electron scattering is dominant below 50~K and short-range scattering is expected. Again in Co$_x$Fe$_{1-x}$Si~\cite{Pshenay2018}, these scattering leads to total electron (hole) transport lifetime at $R$ ($\Gamma$) point to be $\sim~10^{-12}~s$. This value is similar to our experimentally observed $\tau_p$. 

\begin{figure}[]
\centering
\includegraphics[width=0.48\textwidth]{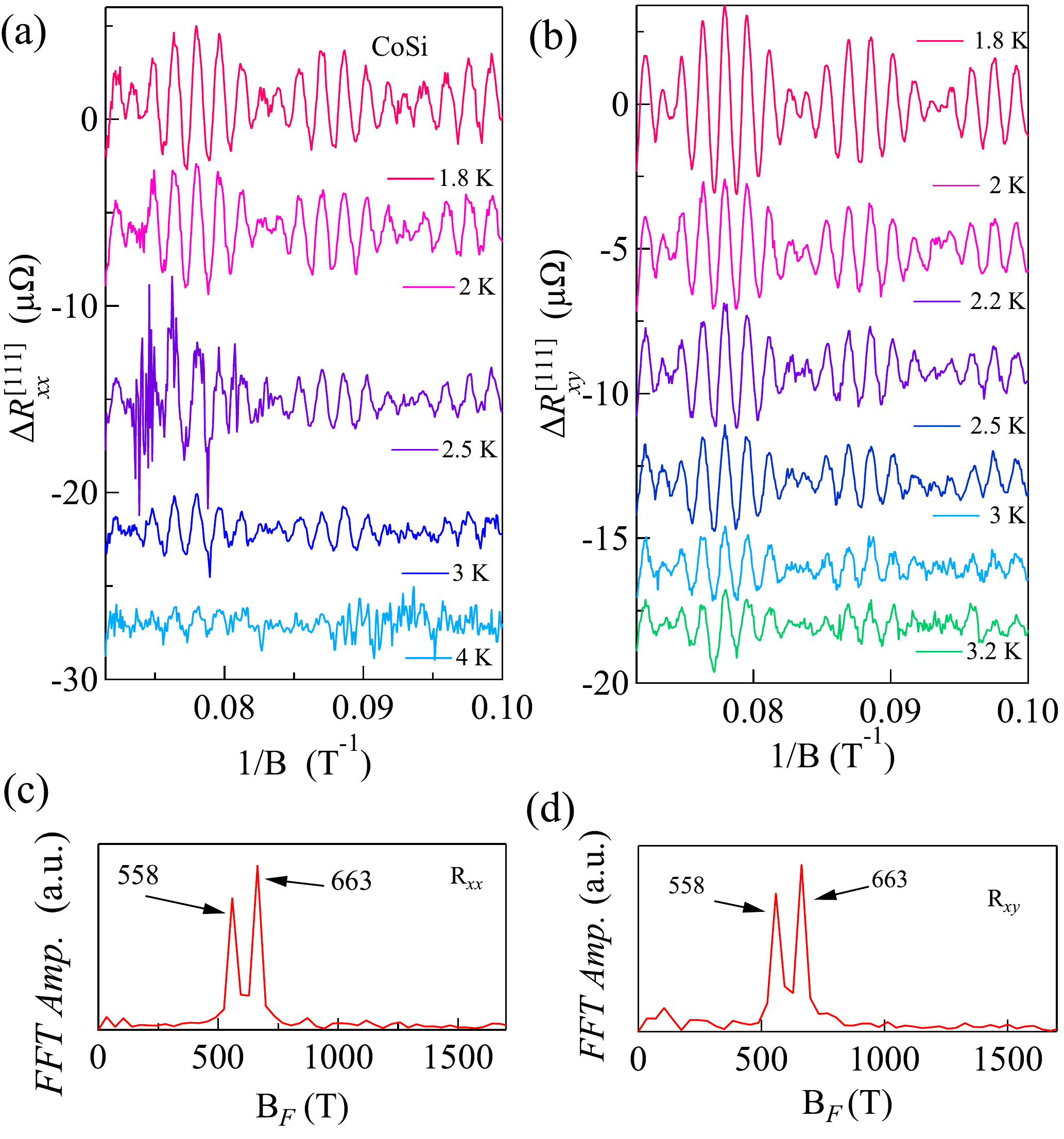}
\caption{(a),(b) SdH oscillation of linear and Hall resistance at different temperature, respectively for $I~\parallel~[111]$ and $B~\parallel~[01$\=1]. (c)-(d) FFT frequencies of the quantum oscillations are shown. Both the data show FFT frequencies of 663~T and 558~T.}
\label{SdH}
\end{figure}

\begin{figure}[]
\centering
\includegraphics[width=0.48\textwidth]{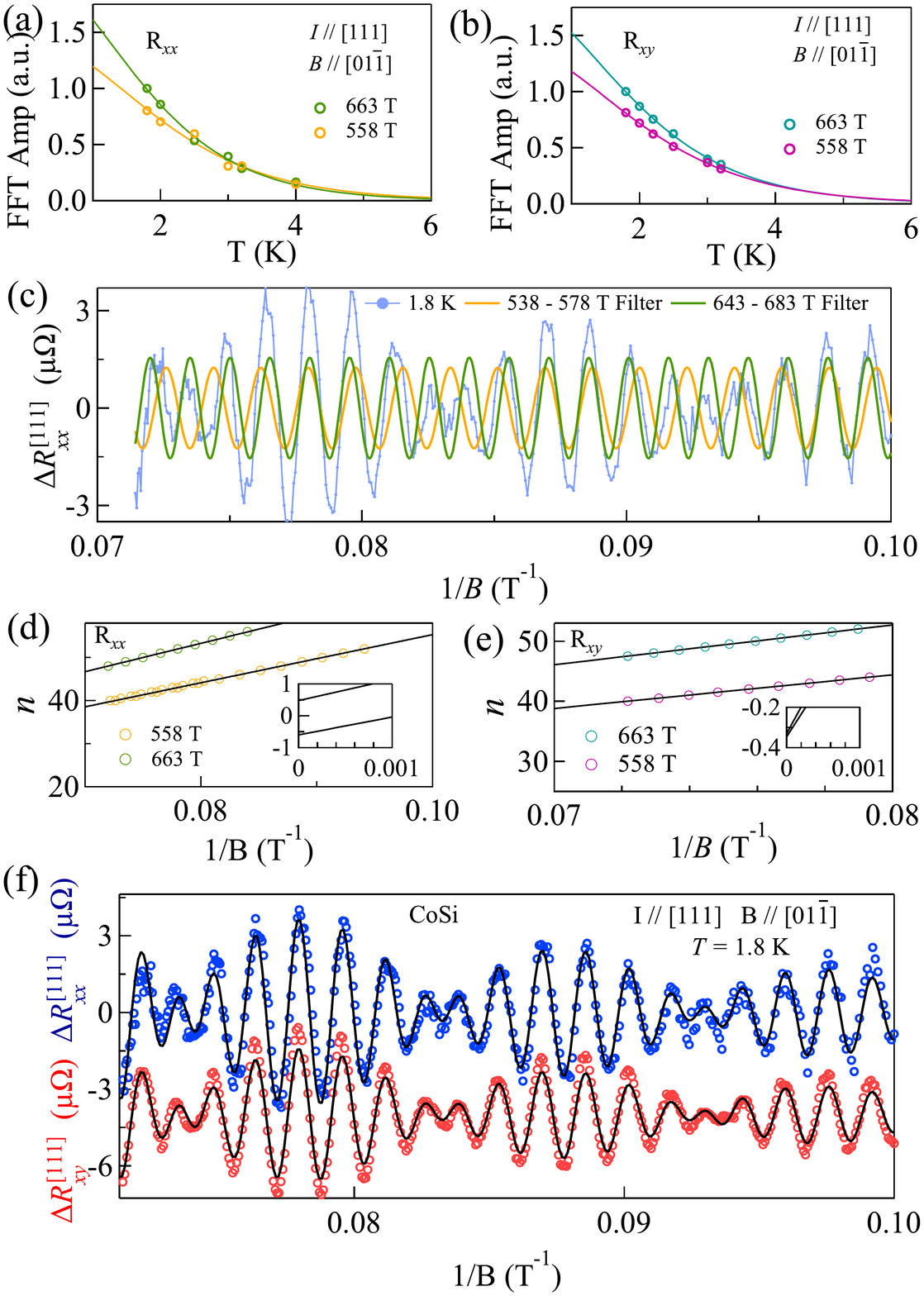}
\caption{(a),(b) $R_{xx}$ and $R_{xy}$ SdH oscillation amplitude as a function of temperature for  $I~\parallel~[111]$ and $B~\parallel~$[01\={1}]. (c) FFT filter used to extract the beating pattern of 663~T and 558~T oscillation at 1.8~K. (d),(e) considering maximum point as $n$ th LL and minimum as ($n+1/2$) th LL, we illustrate the LL fan diagrams for both $R_{xx}$ and $R_{xy}$. (f) LK fit of $\Delta R_{xx}$ and $\Delta R_{xy}$ at 1.8 K. The Dingle temperatures ($T_D$) have been extracted from the LK fitting}
\label{LK}
\end{figure}

\begin{figure}[]
\centering
\includegraphics[width=0.48\textwidth]{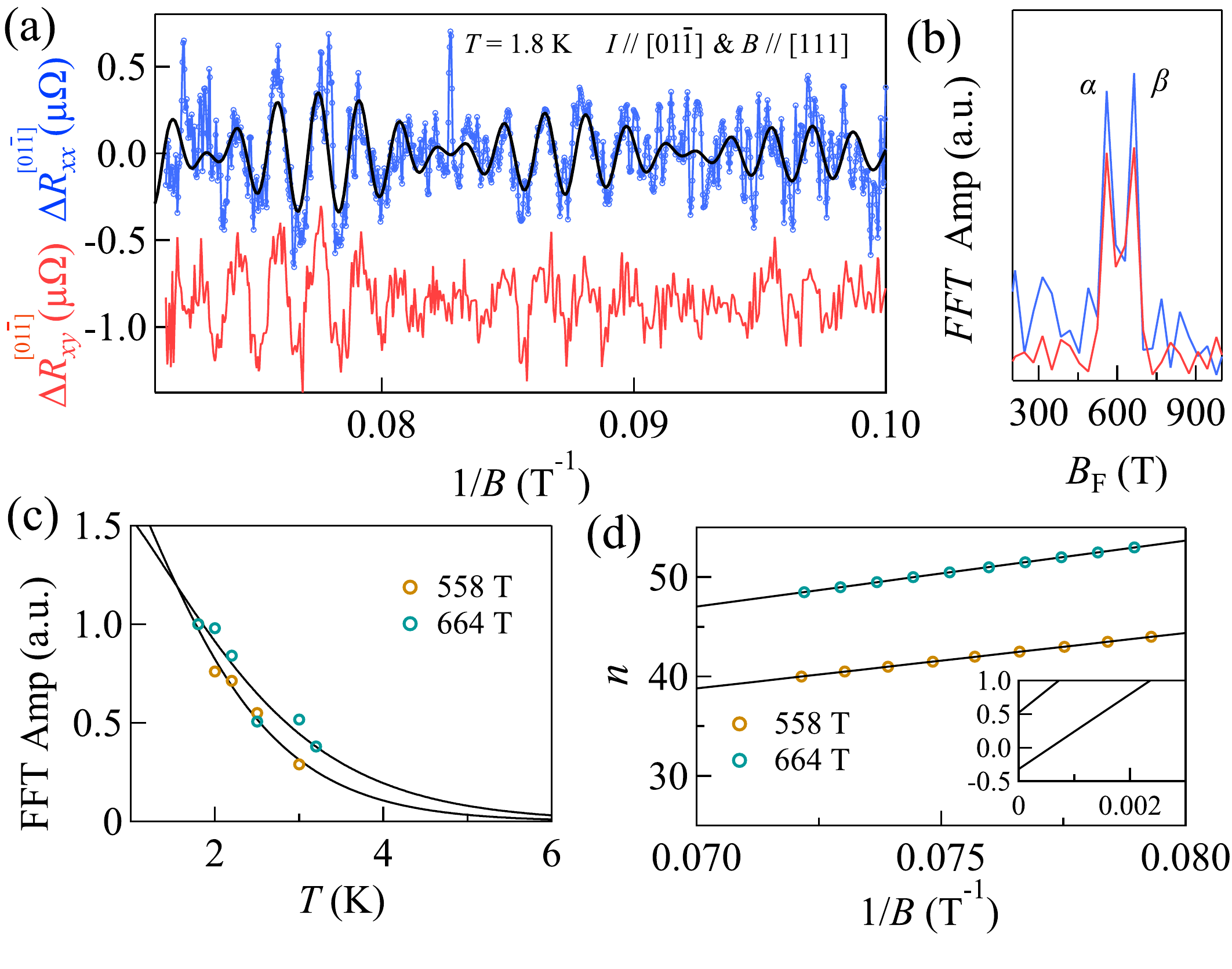}
\caption{(a) Back ground subtracted oscillation $\Delta R_{xx}^{[01\bar{1}]}$ and $\Delta R_{xy}^{[01\bar{1}]}$ for $I~\parallel~$[01\={1}] and $B~\parallel~[111]$. (b) Frequencies of oscillations. (c) calculated temperature dependent oscillation amplitudes of 663~T and 558~T from $\Delta R_{xx}$. (d) LL fan diagram of the oscillations.}
\label{SdH2B}
\end{figure}

\subsection{Shubnikov-de Haas oscillation studies}
\begin{table*}[]
\caption{Calculated quantum parameters from SdH oscillations. Here, $A$ is the extremal cross sectional area, $k_F$ is the Fermi wave vector. $m^*_{xx}$ and $m^*_{xy}$ are effective calculated from linear and Hall oscillations, respectively. $\tau_{q}^{xx}$ and $\tau_{q}^{xy}$ are quantum life time calculated from linear and Hall oscillations, respectively. $n_{SdH}$ is carrier density calculated from SdH.}

\begin{tabular}{cccccccccc} \hline \hline \\
\multicolumn{1}{l}{Exp.} & $I$ \& $B$ & \begin{tabular}[c]{@{}c@{}}Frequency\\ (T)\end{tabular} & \begin{tabular}[c]{@{}c@{}}A\\ (nm$^{-2}$)\end{tabular} & \begin{tabular}[c]{@{}c@{}}$k_{F}$\\ ($10^7~$cm$^{-1}$)\end{tabular} & \begin{tabular}[c]{@{}c@{}}$m_{xx}^*$\\ ($m_e$)\end{tabular} & \begin{tabular}[c]{@{}c@{}}$m_{xy}^*$\\ ($m_e$)\end{tabular} & \begin{tabular}[c]{@{}c@{}}$\tau_{q}^{xx}$\\ ($10^{-13}~s$)\end{tabular} & \begin{tabular}[c]{@{}c@{}}$\tau_{q}^{xy}$\\ ($10^{-13}~s$)\end{tabular} & \begin{tabular}[c]{@{}c@{}}$n_{SdH}$\\ ($10^{23}~$cm$^{-3}$)\end{tabular} \\ \hline \\
\multirow{4}{*}{SdH}  & \multicolumn{1}{l}{\multirow{2}{*}{\begin{tabular}[c]{@{}l@{}}$I~\parallel~[111]$ \&  $B~\parallel~$[01\={1}]\end{tabular}}} & \multicolumn{1}{l}{663 ($\beta$)} & 6.33 & 1.42 & 0.98 & 0.91 & 8.6 & 7.7 & 0.96 \\
 & \multicolumn{1}{l}{} & \multicolumn{1}{l}{558 ($\alpha$)} & 5.33 & 1.3 & 0.86 & 0.85 & 8.7 & 6.9 & 0.74 \\ \\
 & \multirow{2}{*}{\begin{tabular}[c]{@{}c@{}}$I~\parallel~$[01\={1}] \& $B~\parallel~[111]$\end{tabular}} & 664 ($\beta$) & 6.34 & 1.42 & 1.11 & $-$ & 12.9 & $-$ & 0.96 \\
 &  & 559 ($\alpha$) & 5.33 & 1.33 & 1.08 & $-$ & 10 & $-$ & 0.74\\ \hline \hline
\end{tabular}
\label{Table1}
\end{table*}

For $I~\parallel~[111]$, the polynomial background subtracted linear resistance ($\Delta R_{xx}^{[111]}$) and Hall resistance ($\Delta R_{xy}^{[111]}$) are shown in Fig.~\ref{SdH}(a) and (b), respectively. The Fast Fourier Transforms (FFT)  of the oscillations are shown in Fig.~\ref{SdH}(c) and (d)  which depict two frequencies at 558 and 663~T corresponding to $\alpha$ and $\beta$ pockets, respectively. For multiple frequencies, the Lifshitz-Kosevich (LK) equation for SdH oscillation amplitude is given by

\begin{equation}
\label{eq2}
    \Delta R =  \sum_{i=\alpha,\beta}^{} c_i~R_{T}^i~R_{B}^i~\rm{Cos}\left[2\pi\left(\frac{i}{B}+\psi_i\right)\right]
\end{equation}

Here $R_T$ is the thermal damping factor, and $R_B$ is the Field factor. FFT amplitude decreases with increasing temperature and follows Lifshitz-Kosevich thermal damping factor $R_T = (X/sinhX)$, where $X = \frac{2\pi^2k_BTm^*}{\hbar e B}$, $m^*$ is the effective mass, $\hbar$ is the reduced Planck constant and $k_{\rm B}$ is the Boltzmann constant, see Fig.~\ref{LK}(a) and (b). From the fitting, we estimate the effective masses for both linear and Hall transport geometry. At a constant temperature, the SdH oscillation amplitude decreases with decreasing field, and it follows $exp(-\frac{2\pi^2k_BT_Dm^*}{\hbar e B})$. The estimated effective masses ($m^*$) and Dingle temperatures ($T_D$) are listed in Table.~\ref{Table1}. To extract the oscillation corresponding to $\alpha$ and $\beta$ pockets, bandpass filters have been used. Band width filter has been chosen as per the full-width half maxima of the individual oscillation amplitude profiles. Here, in Fig.~\ref{LK}(c), bandwidth of $538-578$~T filter and $643-683$~T filter have been used to extract the beating pattern corresponding 558~T and 663~T, respectively. From the Landau-level (LL) fan diagram of both the filtered oscillation, the total phases have been estimated (shown in Fig.~\ref{LK}(d) and (e)). For $\Delta R_{xx}^{[111]}$ the obtained total phases $\psi_{\alpha} = -0.51$ and $\psi_{\beta} = 0.5$. Whereas, for $\Delta R_{xy}^{[111]}$ the obtained total phases $\psi_{\alpha} = -0.34$ and $\psi_{\beta} = -0.33$. From the extracted parameters  of $F$, $m^*$, $\psi$, the Dingle temperature ($T_{\rm D}$) of the individual frequencies can be calculated from the LK fit of the oscillatory pattern (shown in Fig.~\ref{LK}(f)). From $\Delta R_{xx}$ we obtained $T_D$ as 1.4~K for $\alpha$ and 1.4~K for $\beta$. Similarly, from $\Delta R_{xy}^{[111]}$ obtained $T_D$ is 1.76~K for $\alpha$ and 1.57~K for $\beta$. The quantum transport lifetime ($\tau = \frac{\hbar}{2 \pi k_B T_D}$) can be calculated from $T_{\rm D}$ and the estimated values for the same are given in the Tab.~\ref{Table1}. Similarly, the SdH oscillations have been observed for $I~\parallel~$[01\={1}] and $B~\parallel~[111]$ (see Fig.~\ref{SdH2B}), and the extracted quantum parameters are listed in Tab.~\ref{Table1} . All diagonal and off-diagonal transport lifetime is very much similar. These results suggest that  in bulk 3D CoSi the charge carriers scatter isotropically while measuring linear resistivity or Hall resistivity.

\subsection{de Haas-van Alphan quantum oscillation studies}

\begin{table}[]
\caption{Extracted quantum parameters of dHvA oscillations of $\gamma$ (18~T) frequency. $A$ is extremal cross section area, $m^*$ is effective mass, $T_D$ is Dingle temperature, $\tau$ is quantum lifetime.}
\begin{tabular}{ccccccc} \\ \hline \hline \\
Exp. & $B$ & \begin{tabular}[c]{@{}c@{}}Frequency\\ (T)\end{tabular} & \begin{tabular}[c]{@{}c@{}}A\\ (nm$^{-2}$)\end{tabular} & \begin{tabular}[c]{@{}c@{}}$m^*$\\ ($m_e$)\end{tabular} & \begin{tabular}[c]{@{}c@{}}T$_D$\\ (K)\end{tabular} & \begin{tabular}[c]{@{}c@{}}$\tau$\\ ($10^{-13}$~s)\end{tabular} \\ \hline \\
\multirow{2}{*}{dHvA} & [111] & 18 & 0.17 & 0.16 & 1.87 & 6.5 \\ \\
 & [01\={1}] & 18 & 0.17 & 0.16 & 1 & 12 \\ \hline \hline
\end{tabular}
\label{Table2}
\end{table}

\begin{figure}[]
\centering
\includegraphics[width=0.47\textwidth]{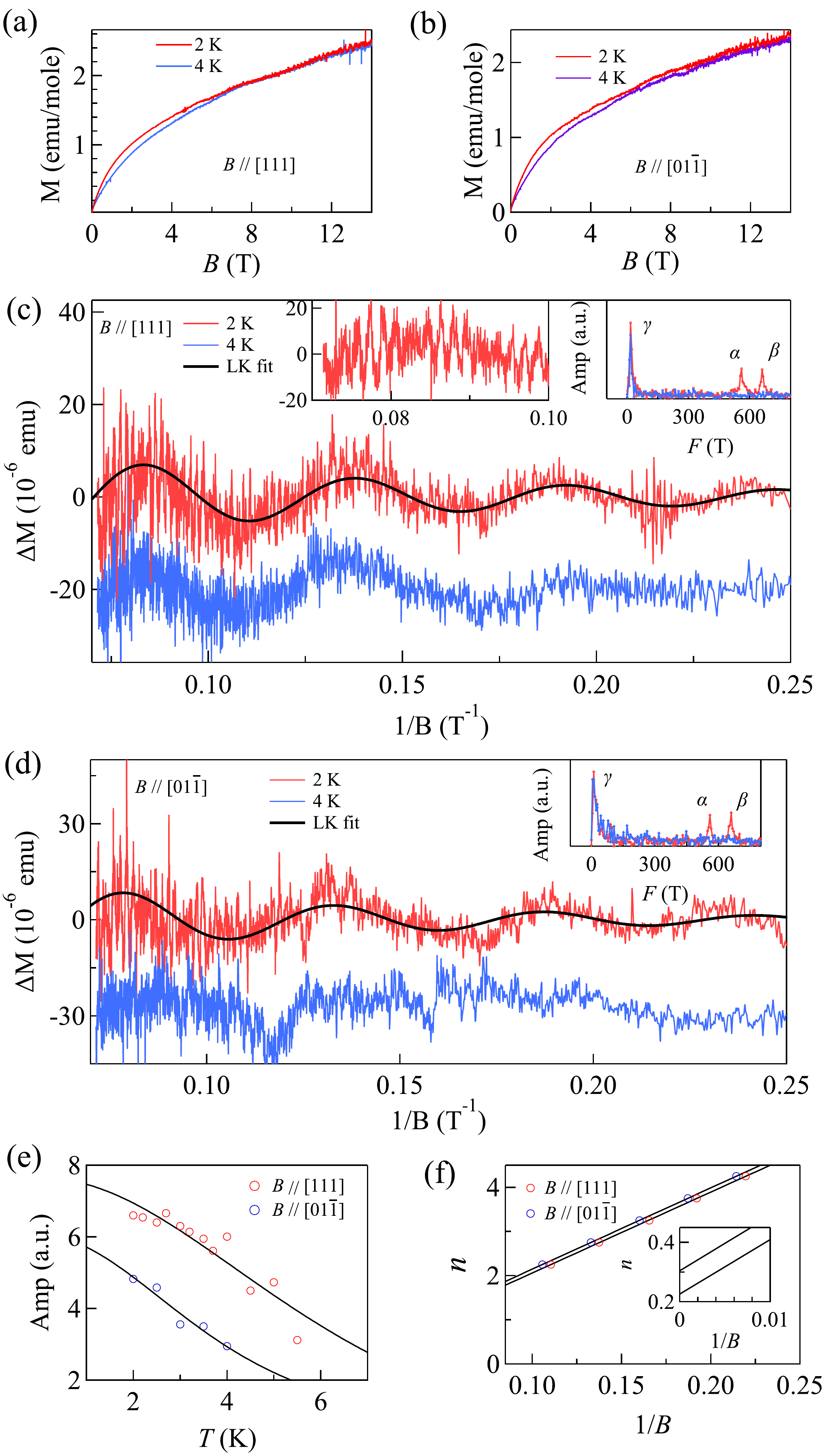}
\caption{(a),(b) Magnetic moment as a function of field for $B~\parallel~[111]$ and $B~\parallel~$[01\={1}] direction, respectively. (c) Background subtracted magnetization ($\Delta M$) as a function of inverse magnetic field ($1/B$) when the magnetic field is applied parallel to $[111]-$ direction. Right inset shows the Fast Fourier transform (FFT) frequencies at 2~K and 5~K. This indicates the absence of $\gamma$ pocket at 5 K where as all $\alpha$, $\beta$ and $\gamma$ are present at 2~K. Left inset shows the zoom version of the $\Delta M$ vs $1/B$ at 2~K. It shows the presence of both $\alpha$ and $\beta$ pocket. (d) $\Delta M$ vs $1/B$ for $B~\parallel~$[01\={1}] at 2 K and 5 K. Inset: FFT frequencies at both temperatures. (e) Temperature dependent dHvA amplitude and mass plot for $\gamma$ pocket (18.4~T). Black like indicate the LK fit. (f) Linear slope of Landau level fan diagram.}
\label{DeltaM}
\end{figure}

The isothermal magnetization data measured at $2$ and $4$~K  are shown in Fig.~\ref{DeltaM}(a) and (b) for different applied field directions.  The magnetizations are very small and remain paramagnetic.  This is in contrast to the Te-flux grown CoSi, where their magnetization was showing a diamagnetic behavior~\cite{PhysRevB.102.115129}.  The nature of magnetic behavior has been under debate~\cite{narozhnyi2013studying}.  The polynomial background-subtracted magnetization ($\Delta M$) measured at $2$ and $4$~K plotted against $1/B$ are shown in Fig.~\ref{DeltaM}(c) and (d).  For $B~\parallel~[111]$, at 2~K, the oscillation has three frequencies, as shown in the inset of Fig.~\ref{DeltaM}(c)  at $18.4$~T ($\gamma$), 558~T ($\alpha$) and 663~T ($\beta$). Inset of Fig.~\ref{DeltaM}(c) shows the zoomed-in high field region which reveals the oscillatory pattern of both $\alpha$ and $\beta$. It is important to mention here that the  oscillation at 4~K has only one frequency and that corresponds to   $\gamma$. Here the FFT pattern in Fig.~\ref{DeltaM}(c) reveals the oscillation frequencies at 2 and 4~K. For $B~\parallel~$[01\={1}], $\Delta M$ vs $1/B$ at 2 and 4~K are shown in Fig.~\ref{DeltaM}(d). The FFT pattern clearly reveals three distinct frequencies as observed for $B~\parallel$~[111] direction.  Similar to $B~\parallel~[111]$, the oscillations have the same frequencies for $B~\parallel~$[01\={1}]. This suggests the isotropic nature of the Fermi surface.  The effective masses and Dingle temperatures are already known from the SdH oscillation studies for $\alpha$ and $\beta$ frequencies. In the magnetization data, we are more interested to look at the $\gamma$ frequency. The oscillation amplitude decreases with increasing temperature and it follows LK formula. A bandpass filter has been used to extract the oscillation corresponding to 18.4~T. Here, the $16-22$~T FFT filter has been used to extract the total phase. So that the LK formula for specific $\gamma$ frequency can be written as:

\begin{equation}
\label{eq3}
    \Delta M =  c~B^{1/2}~R_{T}~R_{B}~\rm{Sin}\left[2\pi\left(\frac{\gamma}{B}+\psi_\gamma\right)\right]
\end{equation}

Figure.~\ref{DeltaM}(e) shows the temperature-dependent amplitude for both $B~\parallel~[111]$ and [01\={1}]. From the mass plot we find the effective masses ($m^*$) are 0.163~$m_e$ and 0.28~$m_e$ for $B~\parallel~[111]$ and [01\={1}], respectively. The $\psi_{\gamma}$ in Eq~\ref{eq3} is the total phase shift in dHvA oscillation corresponding to $\gamma$ pocket. For 3D Fermi surface $\psi$ can be written as $[(\frac{1}{2}-\frac{\phi_B}{2\pi})-\delta]$ where the $\phi_B$ is the Berry phase and $\delta$ assumes as value of $+1/8$ or $-1/8$ for minimum or maximum cross-section area. For topologically trivial band ($\phi_B = 0$), the total phase $\psi$ to be equal to $(1/2\pm1/8)$ = 5/8 or 3/8. Whereas for nontrivial topological band ($\phi_B = \pi$), the total phase $\psi$ should be $\pm1/8$. The frequency $\gamma$ is filtered out from dHvA oscillation. The Landau level fan diagram of the $\gamma$ is shown in Fig.~\ref{DeltaM}(f) where the maxima point are considered as $(n-1/4)$ th Landau level and minima points are $(n+1/4)$. It reveals the $\gamma$ frequency occurs the total phase ($\psi$) to be 0.22 for $B~\parallel~[111]$ and 0.30 $B~\parallel~$[01\={1}]. From the LK fit at 2~K, the Dingle temperature ($T_D$) of $\gamma$ has been extracted for both field directions. In Fig.~\ref{DeltaM}(c) and (d) the black line indicates the LK fit. The obtained $T_D$ are 1.63~K for $B~\parallel~[111]$ and 2.8~K for $B~\parallel~$[01\={1}]. In Ref.~\cite{Sanchez2019}, the angle-resolved photoemission spectroscopy data depicts a tiny hole pockets at $\Gamma$ point which is well separated from the electron pocket at $R$ point. These hole pockets is attributed to the observed  $\gamma$ frequency.

\subsection{Thermodynamic studies}

\begin{figure}[]
\centering
\includegraphics[width=0.4\textwidth]{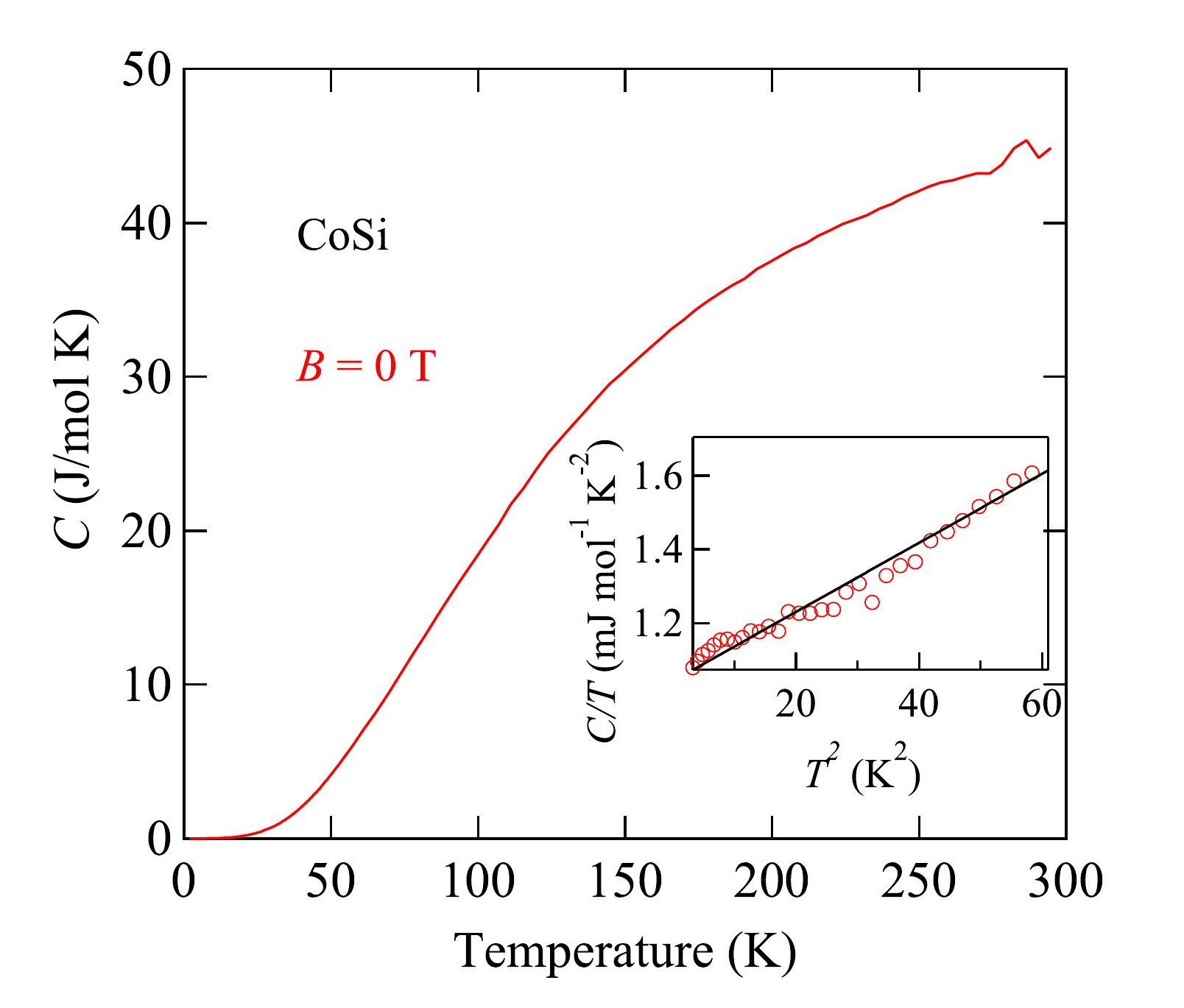}
\caption{Heat capacity ($C$) as function of temperature ($T$) at zero applied magnetic field. Inset: $C/T$ vs $T^2$ and the black solid line indicates the linear fit.}
\label{Cp}
\end{figure}

We have measured the specific heat capacity of CoSi in the temperature range $2 - 300$~K as shown in Fig.~\ref{Cp}.  The $C_{\rm p}$ did not show any anomaly in the measured temperature range as expected for a sample without any magnetic transition.  The $C_{\rm p}$ reaches the Dulong Pet-it limit of  $3nR = 49.89$~J/K mol (where $n$ is the number of atoms in this case it is 2 and $R$ is the molar gas constant) at $300$~K. Inset shows the low-temperature plot of $C/T$~vs.$T^2$.  From the linear fit $C/T = \gamma + \beta T^2$, we have estimated the Sommerfeld coefficient $\gamma$ and the phonon contribution $\beta$ as $1$~mJ K$^{-2}$mol$^{-1}$ and $0.01$~mJ K$^{-4}$ mol$^{-1}$, respectively.  Furthermore, we established density of state at the Fermi level ($\mathscrsfs{D}_\gamma (E_{\rm{F}}) = \frac{3\gamma}{\pi^2 k_B^2}$) to be 0.42 states/eV f.u. From the low-temperature transport, it has already been discussed that the short-range scattering is dominant. The strength of such scattering rate can be realized from the density of states~\cite{Pshenay2018}, which is inversely proportional to the transport lifetime. The expected density of state at the Fermi level, in CoSi, is approximately a few states/eV f.u.~\cite{Pshenay2018}, which is similar to our experimentally observed results.

\subsection{Summary}
In summary, we studied the electrical transport and quantum oscillation in a high-quality CoSi single crystal grown from a stoichiometric melt by the Czochralski method. Here, the anisotropic transport has been measured for $I~\parallel~[111]$ and $I~\parallel~$[01\={1}]. Temperature-dependent longitudinal resistivities indicate the dominant electron-electron interaction. Calculated hole and electron concentration from Hall conductivity at 2~K suggests no carrier compensation. In the present case, the Hall mobility is not very high and it is attributed to the absence of backscattering suppression leading to a lower transport lifetime.  SdH oscillations have been measured for $I~\parallel~[111]$ \& $B~\parallel~$[01\={1}] and $I~\parallel~$[01\={1}] \& $B~\parallel~[111]$. Both directional oscillations result in no such Fermi surface anisotropy. From the quantum oscillation, the extracted quantum parameters indicate the transport lifetime and quantum lifetime are in the same order which can be considered as short-range potential scattering in the system. SdH oscillations in both linear and Hall transport depict two frequencies at 558~T ($\alpha$) and 663~T ($\beta$). dHvA oscillation displays three frequencies 559~T ($\alpha$), 664~T ($\beta$) and 18.4~T ($\gamma$)  at 2~K, whereas the $\alpha$ and $\beta$ are absent for temperature above 4~K. Here, $\alpha$ and $\beta$ are corresponding to two-electron pockets at $R$ point, and $\gamma$ is expected to be tiny hole pockets at $\Gamma$ point.


%

\end{document}